\documentclass[10pt,twocolumn]{article}
\usepackage[a4paper,margin=0.6in]{geometry}
\usepackage{cite}
\usepackage{amsmath,amssymb,amsfonts}
\usepackage{graphicx}
\usepackage{textcomp}
\usepackage{array}
\usepackage{url}
\usepackage{verbatim}
\usepackage{balance}
\usepackage{caption}
\usepackage{subcaption}
\usepackage{xcolor}
\usepackage{gensymb}
\usepackage{algorithm}
\usepackage{afterpage}
\usepackage{dblfloatfix}
\usepackage{todonotes}
\usepackage{multirow}
\usepackage{makecell}
\usepackage{threeparttable}
\usepackage{textcomp}
\usepackage{xcolor}

\usepackage{algpseudocode}
\algtext*{EndIf} 
\algtext*{EndFor} 
\algtext*{EndProcedure} 
\def\BibTeX{{\rm B\kern-.05em{\sc i\kern-.025em b}\kern-.08em
    T\kern-.1667em\lower.7ex\hbox{E}\kern-.125emX}}
\begin{document}

\newcommand{\arxivnotice}{%
\copyright~2025 IEEE. This paper has been accepted for presentation at the Design, Automation and Test in Europe Conference (DATE) 2025. The official version will appear in the DATE 2025 proceedings.
}



\title{ \textbf{A Multi-Stage Potts Machine based on Coupled CMOS Ring Oscillators}  }

\author{Yilmaz Ege Gonul, Baris Taskin \\\textit{Drexel University, Philadelphia, PA, USA,}\\ Email: \{yeg26,bt62\}@drexel.edu}

\date{} 

\maketitle
\footnotetext{
© 2025 IEEE. This paper has been accepted for presentation at the Design, Automation and Test in Europe Conference (DATE) 2025. The official version will appear in the DATE 2025 proceedings.
}

\begin{abstract}

This work presents a multi-stage coupled ring oscillator based Potts machine, designed with phase-shifted Sub-Harmonic-Injection-Locking (SHIL) to represent multivalued Potts spins at different solution stages with oscillator phases. The proposed Potts machine is able to solve a certain class of combinatorial optimization problems that natively require multivalued spins with a divide-and-conquer approach, facilitated through the alternating phase-shifted SHILs acting on the oscillators. The proposed architecture eliminates the need for any external intermediary mappings or usage of external memory, as the influence of SHIL allows oscillators to act as both memory and computation units. Planar 4-coloring problems of sizes up to 2116 nodes are mapped to the proposed architecture. Simulations demonstrate that the proposed Potts machine provides exact solutions for smaller problems (e.g. 49 nodes) and generates solutions reaching up to 97\% accuracy for larger problems (e.g.~2116 nodes).  
\end{abstract}


\section{Introduction}

Solving a certain class of combinatorial optimization problems (COP), called non-deterministic polynomial-time (NP) hard or complete problems, has been historically challenging when using conventional von Neumann type computing systems \cite{COP_exponential}. Given the critical role of solving COPs in various real-world applications, there has been a growing focus on developing accelerated solution methodologies. One of the trending methods for solving COPs is the utilization of Ising machines, which are hardware accelerators specifically targeting COPs. Ising machines solve COPs by mapping the problem constraints onto the Ising spin-glass model \cite{cop_mapping}, and naturally moving to the ground energy states that comprise a solution to the mapped problem. A number of Ising machine implementations built in different technologies (e.g. quantum, optical, oscillator-based, CMOS) exist in the literature~\cite{dwave, CIM, cmos_annealer, wang_oim, prob_fabric, chris_kim_nat, rtwo_ising}. Among these implementations, CMOS ring oscillator based Ising machines~\cite{prob_fabric}, combine the low cost and ease of fabrication of CMOS technology with the non-deterministic, fast nature of physics-based computing.  

Potts model \cite{potts_model} is a generalization of the Ising model which is not restricted to 2 spins like the Ising model but can incorporate multivalued (i.e. $>2$) spins. Potts model is prime for representing a certain class of COPs that natively require multivalued spins such as graph coloring or max-K-cut. Multivalued spins can also represent integers, allowing Potts model to capture a more diverse set of problems including integer optimization problems \cite{potts_model}.  
Several Potts machine implementations are present in the literature, developed with optical \cite{optical_potts}, \cite{coherent_potts}, and hybrid (optical \& digital) \cite{potts_nature} technologies. A CMOS based Potts machine \cite{ICCAD_potts} solving 3-coloring problems through the use of higher order Sub-Harmonic-Injection-Locking (SHIL)  is previously introduced. 

This work proposes a multi-stage CMOS ring oscillator~(ROSC) based Potts machine (MSROPM) architecture that maps and solves COPs requiring multivalued spins by dividing the problem into multiple solution stages. The proposed architecture is able to represent multiple spins through each individual ROSC. This feature is enabled by the binarization of the ROSC phases by phase-shifted SHIL signals into different binary-valued sets of phases at different solution stages. The operation of the MSROPM is through SHILs advancing the oscillators through (i.e. multi-) stages where oscillators act as memory units and computation units. Between these solution stages, no intermediary mapping or caching is necessary as the phase-locked ROSCs are able to preserve their states between solution stages under the influence of SHIL, facilitating in-memory computation. Planar 4-coloring problems of different sizes are mapped to the proposed MSROPM to demonstrate efficiency and scalability.

\section{Background on Ising/Potts Model and Solvers}
\label{sec:ising}
General Ising and Potts model computations are summarized in Section~\ref{sec:theory_ising} and Section~\ref{sec:theory_potts}, respectively. The use of coupled ring oscillators (ROSC) for the Ising/Potts architectures, is reviewed in Section~\ref{sec:litreview_ising}. 

\subsection{General Theory on Ising Model and Machines}
\label{sec:theory_ising}

Ising model, describes the behavior of interacting ferromagnetic spins arranged on a lattice \cite{ising_model}. The energy of the Ising system is described by the Hamiltonian $H(s)$ \cite{ising_model} ignoring the external field term for the scope of this work:
\begin{equation}
  H(s) = \sum_{i,j} J_{ij}s_is_j,
\end{equation}
where $s_i \in \{-1,+1\}$ is a binary valued spin on the $i^{th}$ node on the model and $J_{ij}$ is the coupling strength between the connected spins $s_i$ and $s_j$.

In an oscillator-based Ising machine~(OIM)~\cite{wang_oim},
two distinct phase values of $\theta_{s_i}$, and  $\theta_{s_j}$ ideally $180 \degree$ apart (e.g. 0 \degree, 180\degree), represent the binary spins in the Ising hamiltonian $H(\theta_{s})$ :
\begin{equation}
  H(\theta_{s}) = \sum_{i,j} J_{ij}cos(\theta_{s_i} - \theta_{s_j})
\end{equation}
The interaction strength $J_{ij}$ in an OIM, is arbitrated through the strength of the coupling between the oscillators. 

\subsection{General Theory on Potts Model and Machines}
\label{sec:theory_potts}

Potts Model~\cite{potts_model} is the generalization of the Ising model, describing the interactions of  multivalued (e.g. $ > 2$) spins arranged on a lattice.  
Consequently, Potts machines are prime for solving COPs that cannot be modeled natively with 2 spins (i.e. with an Ising machine) more effectively.  
The standard Potts model Hamiltonian $H_{Potts}$ is: 
\begin{equation}
  H_{Potts} = \sum_{i,j} J_{ij} \delta (s_i,s_j),
\end{equation}
where $s_i \in \{0,1,2, ...., N-1\}$  is an N-valued spin on the
$\textit{i}^{th}$ node of the model, and $J_{ij}$ is the interaction strength between the spins $s_i$ and $s_j$.

The vector Potts Hamiltonian, also modelling a phase-interaction based Potts model such as in the case of a theoretical oscillator based Potts machine (OPM)~\cite{seal_potts}, $H_{v_{Potts}}$ is:

\begin{equation}
\label{eq:vector_potts}
  H_{v_{Potts}}(\theta_{s}) = \sum_{i,j} J_{ij} cos(\theta_{s_i} - \theta_{s_j}),
\end{equation}
where $\theta_{s_i} - \theta_{s_j}$  is the phase difference between the $\textit{i}^{th}$ and $\textit{j}^{th}$ oscillator. For an N-phase Potts Machine, the oscillator phases $\theta_{s_i}$ are locked at N distinct possible phases equally spaced within the range $[0,2 \pi]$, where $\theta_{s_i} =  \frac{2\pi s_i}{ N}$ and $s_i \in \{ 0, 1, 2, ... N - 1\} $.

The inefficency of the Ising model for a certain class of COPs can be well-understood looking at the Ising formulation of the graph coloring problem. Coloring a given graph $G(V,E)$ having n vertices, using N-colors is phrased~\cite{many_ising} using the Ising model as:
\begin{equation}
  H(s) = J \sum_{i}^{n} (1- \sum_{k=1}^{N} s_{ik})^2  + J \sum_{(i,j) \in E} \sum_{k=1}^{N}   s_{ik}  s_{jk}
\end{equation}
where $s_{ik}$ is a binary-valued spin which is 1 if the $i^{th}$ vertex is colored by the color $k$. In this scheme, $N$ distinct spins (binary-valued) are required for each one of the $n$ vertices of the graph to represent all possible colors, with in total $n \cdot N$ spins. Instead, when Potts model is used to represent the N-coloring problem (the native formulation of N-coloring \cite{many_ising}), a single multivalued spin can assume arbitrarily many values that correspond to different colors a vertex can be assigned, leading to a representation with only $N$ spins (1 spin per vertex).  

\begin{figure}[t]
    \centering
    \includegraphics[width=\linewidth]{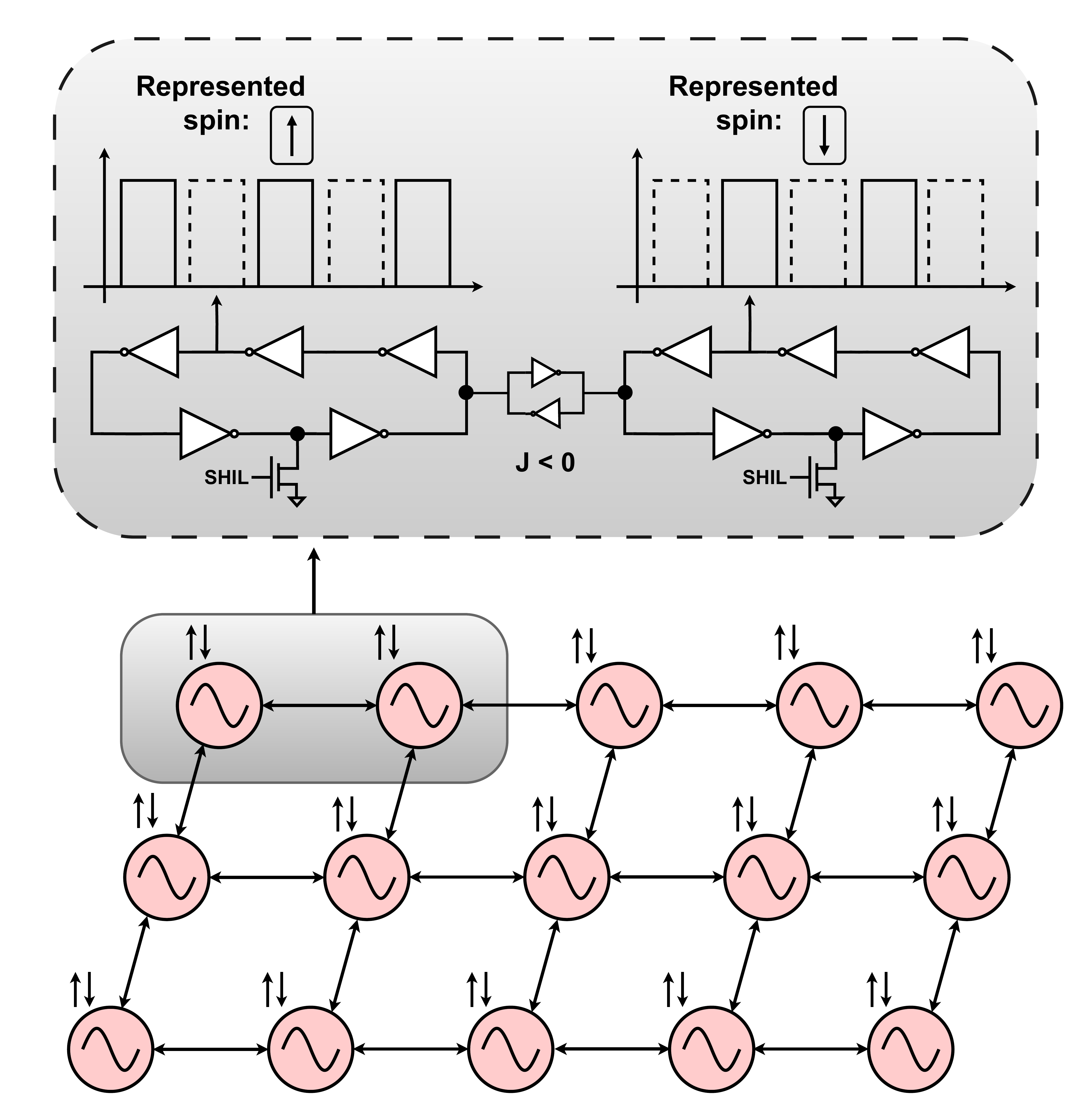}
     \caption{A coupled oscillator array representing an Ising lattice implemented with negatively coupled ROSCs }
    \label{fig:rosc_array}
\end{figure}

\begin{figure*}[t]
    \centering
    \includegraphics[width=\linewidth]{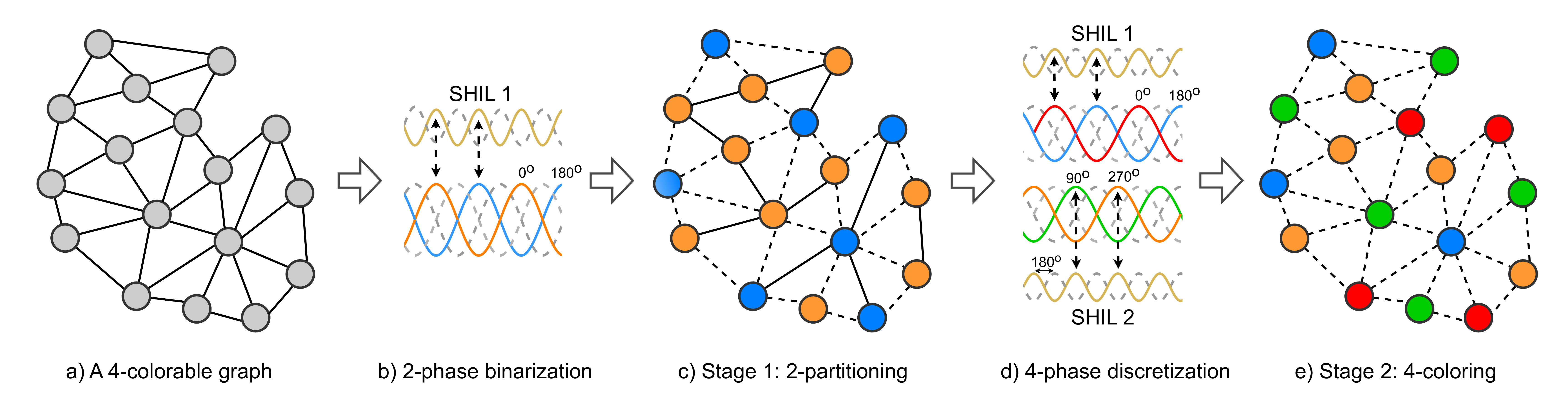}
     \caption{An example 4-colorable graph 4-colored in 2 stages and the corresponding phase-shifted SHILs achieving 4-phase discretization}
    \label{fig:iterative_ops}
\end{figure*}

\subsection{Review on Coupled ROSC based Ising and Potts machines}
\label{sec:litreview_ising}

In the recent literature ROSC based Ising~\cite{prob_fabric}, \cite{chris_kim_nat}, \cite{chris_kim_560} or Potts~\cite{ICCAD_potts} machine~(ROIM and ROPM) implementations using coupled ROSCs solving max-cut and 3-coloring problems are presented. Networks of coupled ROSCs, similar to other types of oscillators \cite{wang_oim}, represent interacting spins through injection locking \cite{injection_lock} acting as coupling between coupled ROSCs \cite{prob_fabric}. Ising or Potts spins are represented by the phase of the individual oscillators. 
The couplings (realized by a current conducting medium such as pass transistors or back-to-back (B2B) inverters \cite{prob_fabric}, \cite{chris_kim_nat}) facilitate the phase interactions through the natural energy minimization of the system. As depicted in Figure~\ref{fig:rosc_array}, negative coupling (by B2B inverters which is an inverting coupling medium) of ROSCs pushes the coupled ROSCs out of phase and vice-versa for positive coupling \cite{prob_fabric}. When a ROIM reaches a ground state, it is possible for the oscillator phases to not end up at one of the two phases corresponding to the spin values. As a remedy, SHIL, an external perturbation carrying twice the frequency of the oscillator, is utilized to binarize the oscillator phases \cite{wang_oim}. The ROPM in \cite{ICCAD_potts} uses a higher-order SHIL (3-SHIL) to discretize the oscillator phases at 3 distinct phases to represent 3-valued Potts spins. In a ROIM or ROPM, SHIL is introduced by a current-inducing medium such as an NMOS or PMOS transistor~\cite{prob_fabric} as depicted in Figure~\ref{fig:rosc_array}.

Implementation of a ROIM or ROPM requires the tuning of several design parameters such as the strength of couplings and the strength of the SHIL \cite{prob_fabric}. Although stronger couplings allow the system to converge to a ground state (i.e. a solution) faster, coupling strength above a certain threshold can halt the oscillation of the ROSCs. Similarly, SHIL injection below a certain level of strength cannot discretize the ROSC phases and deforms the ROSC waveforms when exceeds a certain level of strength. Although, ideally, ROIMs implemented in all-to-all topology can map graphs of any connectivity, sparser topologies such as hexagonal~\cite{prob_fabric}, or king's graph~\cite{chris_kim_nat} using nearest-neighbor coupling are preferred.

\section{Multi-Stage Coupled ROSC based Potts Machine (MSROPM) Design}
\label{sec:design}

The proposed MSROPM architecture is introduced on the graph coloring problem, following from the general theory of Potts machines presented in Section~\ref{sec:ising}. The proposed divide-and-color approach is introduced in Section~\ref{sec:mapping}. The multi-stage operation and the enabling design components are presented in Section~\ref{sec:multi_stage_ops}.  The circuit implementation details of the MSROPM are presented in Section~\ref{sec:circuit_design}.

\subsection{Graph Coloring and Divide-and-Color}
\label{sec:mapping}

Graph coloring, also referred to as K-coloring, is a well-established COP that involves, given a simple graph $G(V,E)$, assigning one of K colors to each vertex such that no two adjacent vertices share the same color.  
One method to accelerate the solution process of this NP-complete problem has been to divide the graph into smaller parts to solve separate, smaller coloring problems to be later merged back. Some of the numerous software based divide-and-color approaches include \cite{div_conq1,div_conq2 ,div_conq_recursive}. A previous coherent Potts machine (CPM) implementation combining coherent Ising machine with digital feedback~\cite{potts_nature} also leverages this method to solve graph coloring problems. The proposed MSROPM, solves the 4-coloring problem similarly by dividing the problem into 2 stages of max-cut problems. The Potts formulation in Equation~\ref{eq:vector_potts} with $N=4$ is realized by the usage of 2 phase-shifted SHILs, forcing the binarization of the ROSC phases into 2 different sets of target phases at 2 different stages to ultimately obtain 4 equally spaced distinct phases (i.e. Potts spins). Figure~\ref{fig:iterative_ops} demonstrates an example 4-coloring problem solved using the proposed divide-and-color method by first solving max-cut on the initial graph, then solving max-cut again on the partitioned graphs. This methodology can be extended to solve coloring problems with more colors, by adding more solution stages, and using more SHILs.

\begin{figure*}[t]
    \centering
    \includegraphics[width=\linewidth]{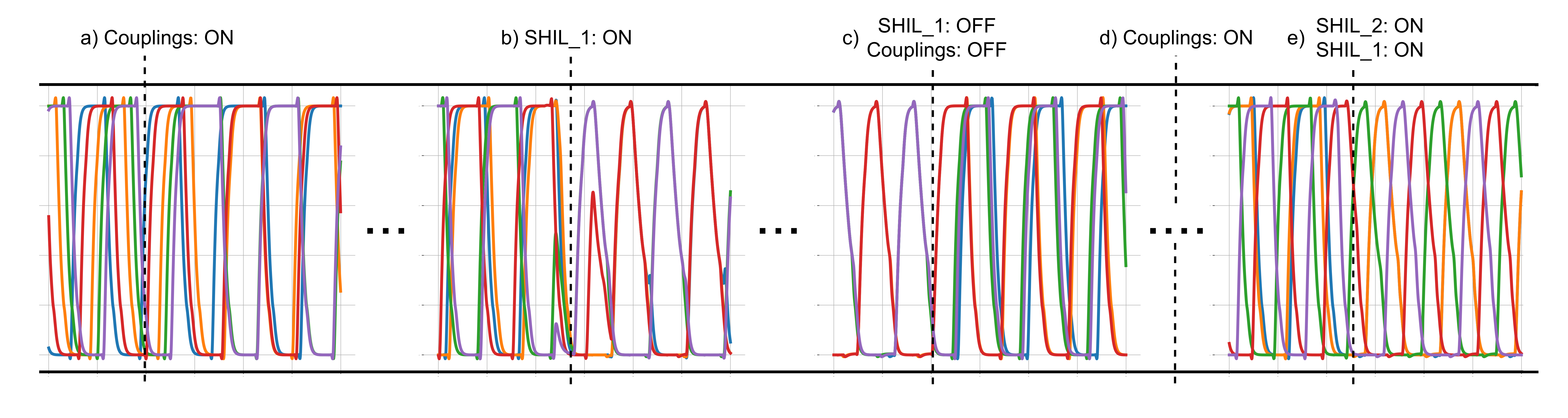}
     \caption{Simulated ROSC waveforms showing the progression of the MSROPM computation cycles and how multi-phase stability is achieved with 2 SHILs. MSROPM operation: a)~activation of couplings, b)~the injection of SHIL\_1 to partition the graph into 2 sets. c)Deactivation of~SHIL\_1 and couplings to allow reinitialization of phases to later d)~turn on the couplings of the 2 independent sets. At the end, e)The injections of ~SHIL\_1 or SHIL\_2 at ROSCs to achieve 4-phase stability}
    \label{fig:waveform}
\end{figure*}

\subsection{Multi-Stage Operation of the Proposed MSROPM}
\label{sec:multi_stage_ops}

The proposed MSROPM utilizes SHIL to achieve three properties. First, MSROPM achieves Potts behavior in multiple stages by using phase-shifted SHILs. Second, SHIL allows the MSROPM to compute-in-memory, thanks to a property of SHIL that allows oscillators to function as both compute and memory units \cite{phlogon}. Third, SHIL clocks the operation of the MSROPM, where the transitions between the stages of computation occur with alternating SHILs. 

A property of SHIL is that the two phases that result from the SHIL binarization, w.r.t. an arbitrary reference point, depend on the phase of the SHIL signal itself w.r.t the same reference point. As Figure~\ref{fig:iterative_ops}(d) exemplifies, whereas SHIL\_1 allows the oscillator phase to stabilize at $0 \degree$ or $180 \degree$, SHIL\_2, that is $180 \degree$ out of phase with SHIL\_1, allows binarization at $90 \degree$ or $270 \degree$. Ultimately, ROSCs in the system end up at one of these four phases corresponding to the 4 colors in the 4-coloring problem. This scheme can be extended to capture an arbitrary number of different stable phases (Potts spins) and solve graph coloring with more colors by increasing the number of SHILs that are shifted in phase.

The compute in memory feature of MSROPM is enabled through the SHILs enabling oscillators to act both as compute and memory units. The CPM in \cite{potts_nature} uses digital components for both forward and backward computations at each stage to solve graph-coloring in multiple stages. Practically, any Ising machine can be used to solve graph coloring in multiple stages of divide-and-color, by reprogramming and remapping the system at each stage and saving the system state in memory between stages. Trying to achieve Potts model computation in this scheme would suffer from the von Neumann bottleneck where the speed advantages of the Ising machine can be nullified due to the time spent on saving and loading the system state. The proposed MSROPM performs the multi-stage operation by computing-in-memory. During the compute stage, coupled ROSCs, free of SHIL influence, naturally move (i.e. self-anneal) towards ground states where phases gradually change. Under the influence of SHIL, oscillator phases are locked, acting as latches (i.e. memory), preserving the current phase until the SHIL is removed.

The third feature of MSROPM thanks to SHILs is the clocking of multiple solution stages through multiple cycles of computation. Similar to the operation of ROSC-based Ising machines \cite{prob_fabric}, \cite{chris_kim_nat}, MSROPM starts with a random initialization of ROSCs before couplings are turned on. After a predetermined amount of time (decided experimentally) first SHIL\_1 is injected to the ROSCs, binarizing the phases into one of the two $180~\degree$ apart possible phases ( $0~\degree$ and $180~\degree$ for convenience). Shortly after SHIL injection, the couplings between opposite-phase ROSCs are turned off, partitioning the circuit into 2 independent sub-graphs. Meanwhile, SHIL is removed from all ROSCs, allowing them to act freely under coupling once more. However, the selected SHIL of the ROSCs having $180~\degree$ phases (i.e. out of phase w.r.t the REF signal) is now toggled. After a predetermined amount of time, two different SHILs (SHIL\_1 on the previously $0~\degree$ set and SHIL\_2 on the $180~\degree$ set) are injected on the two partitions of the circuit, bringing out a final state of phases that are stabilized at 4 equally spaced phases. Figure~\ref{fig:waveform} demonstrates simulated waveforms at each computation stage of the MSROPM, first 2-partitioning the graph (by max-cut) through 2 stable phases, then 4-coloring the graph (by 2 separate max-cuts) with 4 stable phases. In contrast to the ROPM in \cite{ICCAD_potts} that solves 3-coloring problems by enabling the representation all 3 allowed spins through individual ROSCs at a single stage, the proposed MSROPM divides the phase discretization into multiple stages, using only 2-SHILs.



\begin{figure}[]
    \centering
    \includegraphics[width=\linewidth]{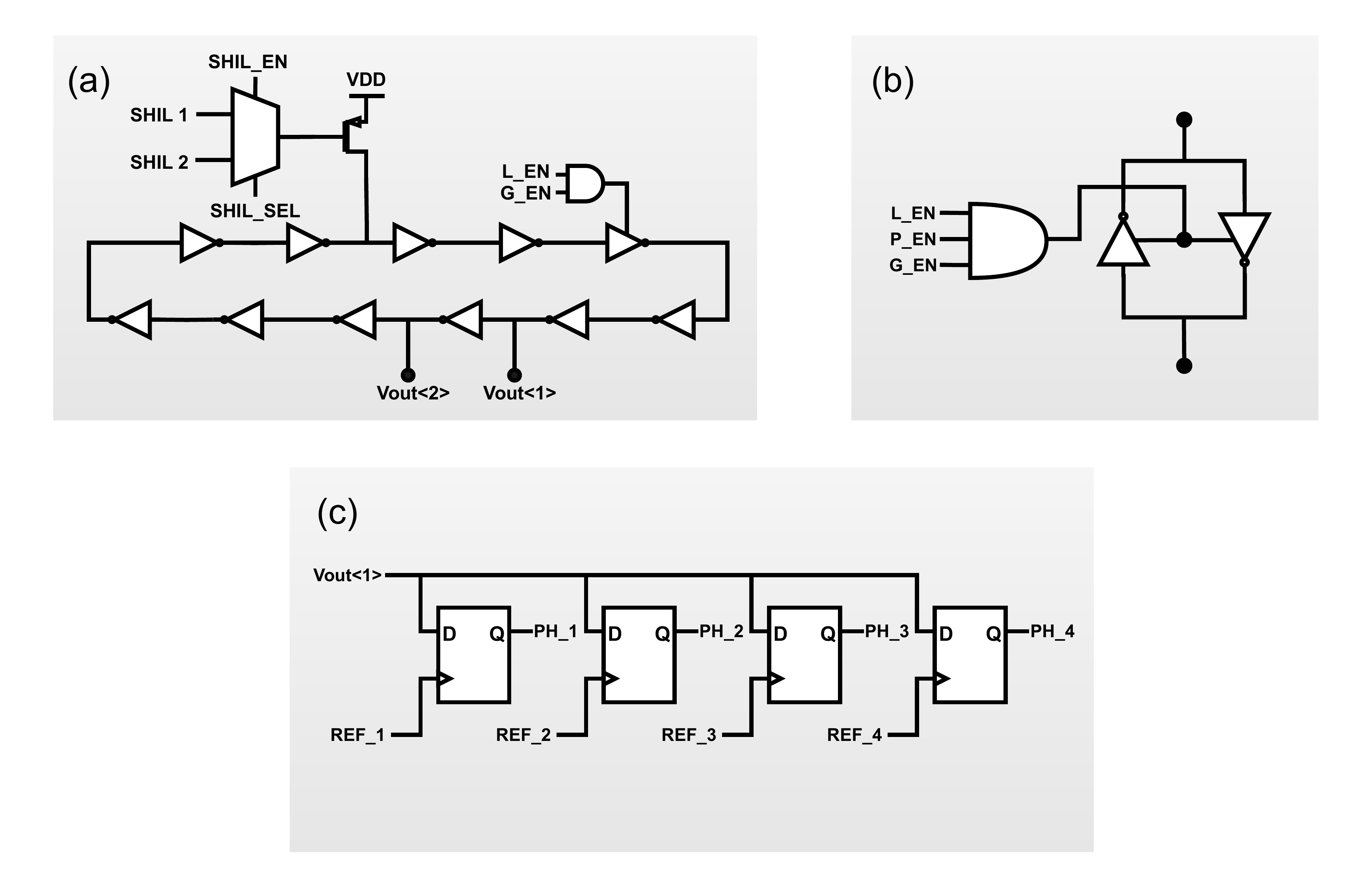}
     \caption{MSROPM experiment setup: a) ROSC block containing 11 inverters and two phase-shifted SHILs alternated through a MUX b) Coupling block including gated B2B c) Phase sampling block }
    \label{fig:blocks}
\end{figure}

\begin{figure*}[t]
     \centering
     \begin{subfigure}[]{0.32\linewidth}
         \includegraphics[width=\linewidth]{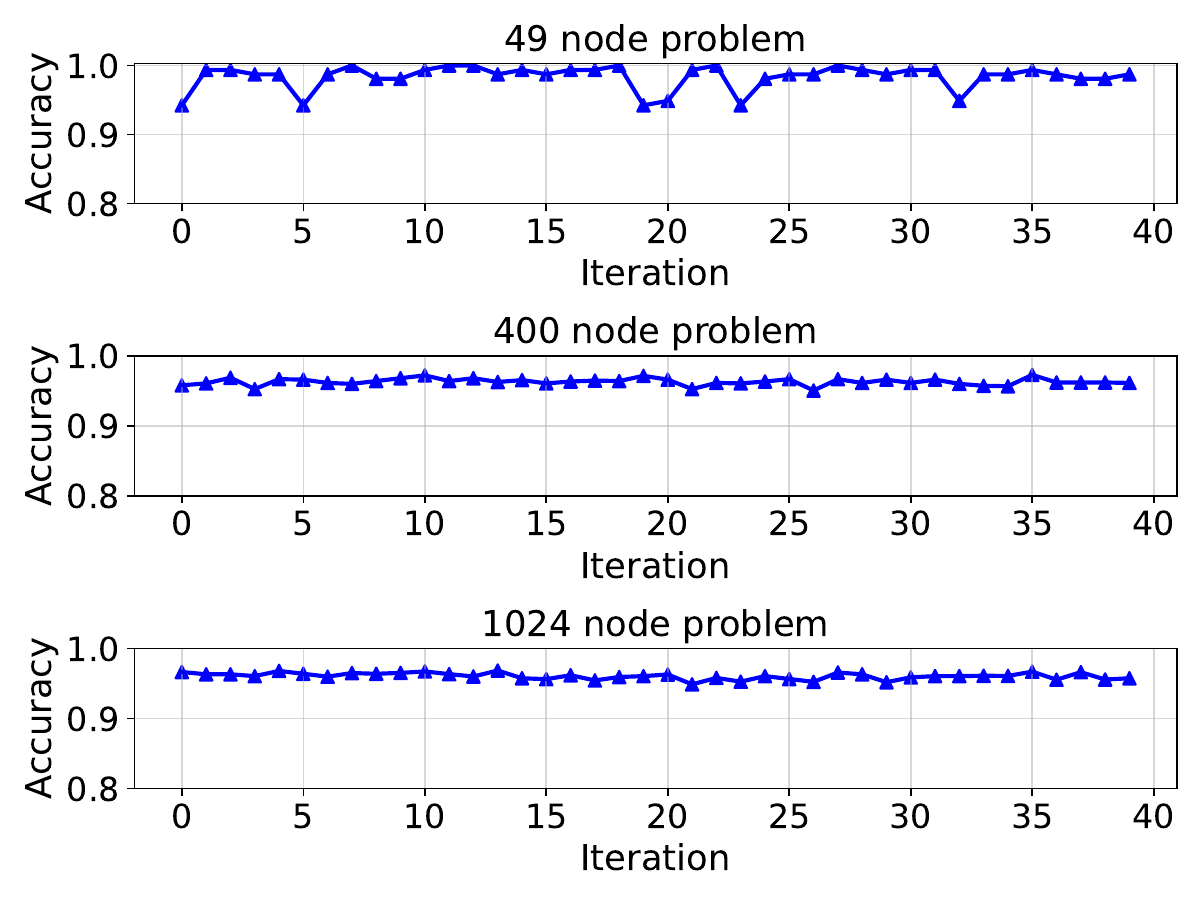}
         \caption{$2^{nd}$ stage 4-coloring accuracy}
    \label{fig:monte_carlo_single}
     \end{subfigure}
     \begin{subfigure}[]{0.32\linewidth}
         \includegraphics[width=\linewidth]{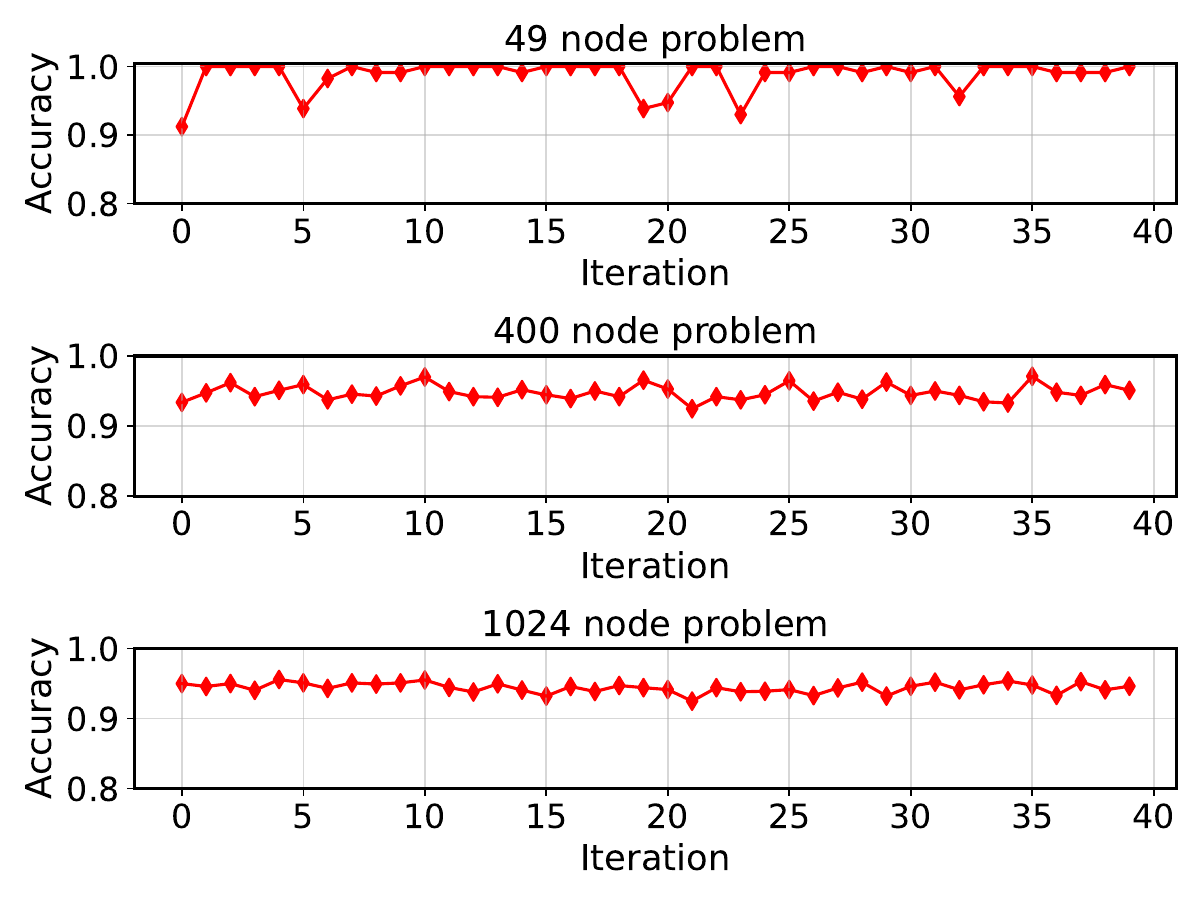}
         \caption{$1^{st}$ stage max-cut accuracy}
         \label{fig:comp_monte_carlo}
     \end{subfigure}
     \begin{subfigure}[]{0.32\linewidth}
         \includegraphics[width=\linewidth]{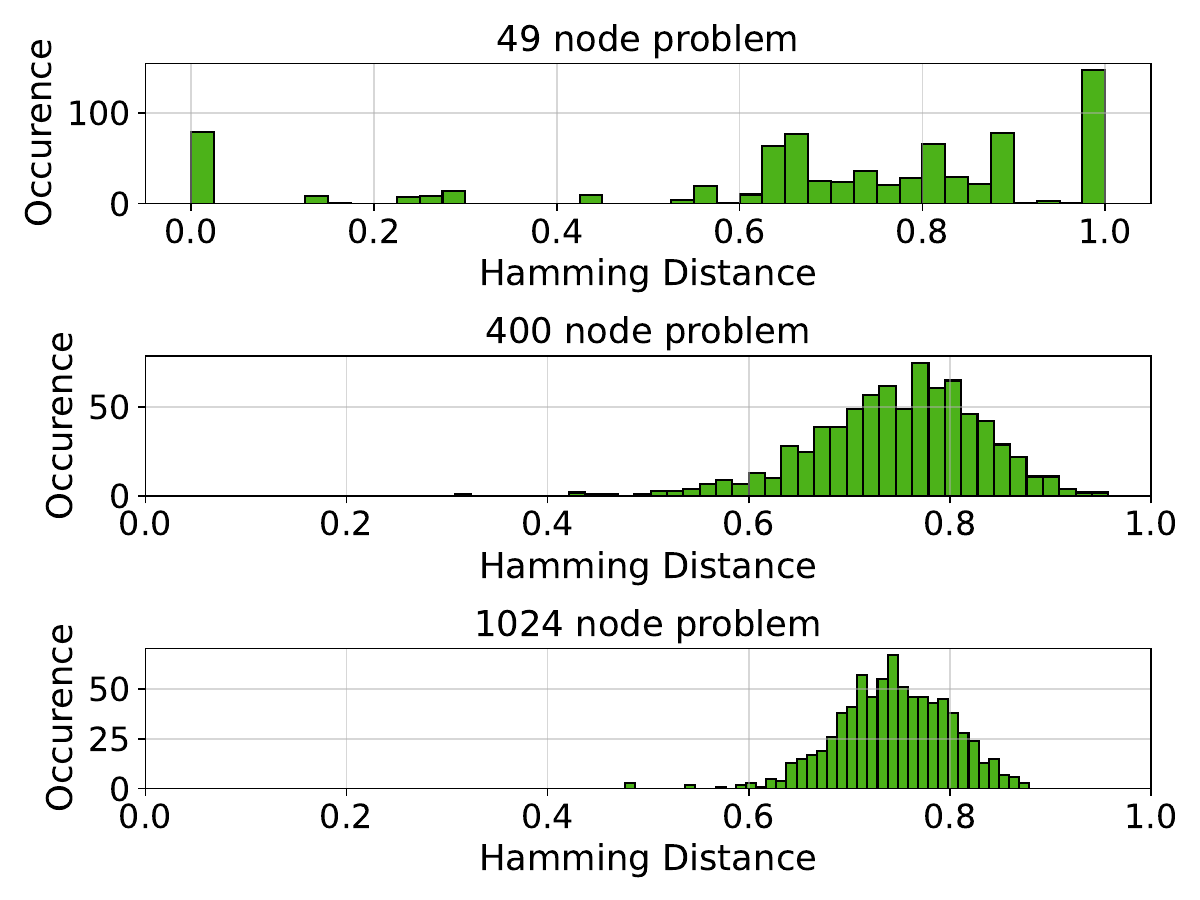}
         \caption{Hamming distance of solutions}
         \label{fig:multiple_instances}
     \end{subfigure}
     \caption{a) 4-Coloring accuracy obtained from 40 iterations (i.e. repeated runs) on 3 different size problems b) max-cut accuracy obtained in the $1^{st}$ stage of computation c) hamming distances of the solutions}
    \label{fig:plots}
\end{figure*}


\subsection{MSROPM Circuit Design}
\label{sec:circuit_design}
An 11-stage ROSC is designed to operate at 1.3 GHz frequency. The inverters are sized with $4:1$ PMOS to NMOS width ratio to enable $2^{nd}$ order SHIL susceptibility of the ROSCs~\cite{analysis_shil}. The couplings between oscillators are through B2B inverters. Both ROSC and B2B inverter units are controlled by global and local enable signals G\_EN and L\_EN. Global signals toggle the ROSCs and B2Bs all at once. Local signals toggle ROSCs and B2Bs individually and are used to map problems to the circuit. Another control signal P\_EN selectively toggles couplings based on phase read-outs to enable partitioning. In an example scheme, when an oscillator locked at the phase $180 \degree$ during the first SHIL as shown in  Figure~\ref{fig:waveform}(b), P\_EN signal of the couplings with neighboring oscillators 
is set 0/1 depending on their phases being $0 \degree$/$180 \degree$ cutting off the coupling between different-phased oscillators ($1^{st}$  stage partitioning).

The SHIL is introduced to each ROSC through a PMOS transistor as depicted in Figure~\ref{fig:blocks}(a). The alternation between the two SHILs is performed via a MUX where the control signal SHIL\_EN enables and disables the SHIL injection and the control signal SHIL\_SEL selects the active SHIL. The value of SHIL\_SEL is determined by the phase readout during the first SHIL (similar to P\_EN). SHIL\_EN is effectively used for disabling all SHILs where self-annealing of coupled ROSCs must occur such as in Figure~\ref{fig:waveform}(a) and Figure~\ref{fig:waveform}(c). For the discretization of phases via the SHILs, as mentioned in Section~\ref{sec:litreview_ising}, the injection strength of the SHIL is critical. A weak SHIL does not discretize the phases with precision, whereas a strong SHIL deforms the waveforms preventing phase readability.  

The phase-readout circuit utilizes reference signals and D-flip-flops (DFFs) to sample the ROSC phases. Under the influence of SHIL, locked ROSC phases are absolute with respect to a certain reference signal, allowing phase sampling solely using DFFs. The rising edges of the reference signals are located at points corresponding to the different phases that correspond to the Potts spins. In this scheme, only one of the 4 (for 4-coloring) DFFs becomes high at a time, capturing the ROSC phase with the resolution of 4 phases. Solving coloring problems with more colors would require increased phase readout resolution hence more DFFs.

\section{ Experimental Results}
\label{sec:simulations}

Simulations are performed on four custom-built MSROPM implementations to investigate the accuracy, speed, and power of the architecture against different 4-coloring problems. The circuits are designed using a 65~nm technology node at 1~V operation. Due to the lack of commonly accepted benchmark problems, custom 4-coloring problems in King's graph topology are generated in different sizes. The SHIL signal and reference signals are simplified as external square waves in simulations. In practice, multi-phase on-chip clock generators~\cite{rotasyn} can be used to generate both SHIL and reference signals. 
To randomly initialize the ROSC phases in simulations, ROSCs are initially turned on at random time instances and set free for an empirically predetermined amount of time to randomly drift apart from each other through jitter \cite{chris_kim_nat}. 




The computation principle of the proposed MSROPM, like other Ising/Potts machines, is probabilistic in nature, i.e. different runs end up with different solutions caused by changing initial conditions (i.e. initial phases of the ROSCs). It is a common observation that probabilistic computing methods do not always reach the optimum solution, instead, they often yield quasi-optimum solutions. Therefore, running multiple iterations for a problem increases the chance of achieving a better solution. In practice, an Ising/Potts solver is typically run multiple times, with the best solution among the iterations being selected as the final solution.

The quality of results is assessed by counting the number of edges in the graph that adhere to the coloring rule for the nodes to which the edges connect. The normalized number of correctly colored neighbors indicates how closely the generated solution approximates the actual solution (global minimum). This accuracy metric is analogous to the normalized Hamiltonian of the solution state relative to the Hamiltonian of the exact solution. Exact solutions of the problems are computed using a generic SAT solver, which serves as the baseline for evaluating accuracy.

\begin{table}[]
\centering
\caption{Statistics from the simulations}
\label{table:sol_size}
\small
\resizebox{\columnwidth}{!}{%
\begin{tabular}{ || c |c |c | c|c ||} 
\hline
  Graph size: & \textbf{49-node} & \textbf{400-node}  & \textbf{1024-node} & \textbf{2116-node} \\ 
\hline
\hline
Search space & $ 4^{49}$ & $ 4^{400}$  & $ 4^{1024}$ & $4^{2116}$ \\ 
\hline
Iterations & 40 &40 &40 & 40 \\
\hline
Average power & 9.4 mW & 60.3 mW & 146.1 mW & 283.4 mW  \\ 
\hline
Top accuracy & 1.00 & 0.98 & 0.97 & 0.97 \\ 
\hline
\end{tabular}%
}
\end{table}

\begin{table*}[th]
\centering
\caption{Comparison with prior work}
\label{table:comparison}
\fontsize{7.5pt}{10pt}\selectfont
 \begin{threeparttable}
\begin{tabular}{| c |c| c| c |c |c | c|c|} 
\hline
 & \textbf{This work} & \textbf{\cite{ICCAD_potts}} & \textbf{\cite{potts_nature}} & \textbf{\cite{optical_potts}} & \textbf{\cite{rtwo_ising}} & \textbf{\cite{chris_kim_nat}}   \\ 
\hline
 \textbf{Solver type} & Potts & Potts  & Potts & Potts & Ising & Ising \\ 
\hline
 \textbf{Solved COP} & 4-coloring & 3-coloring  & 4-coloring & 3-coloring  & Max-Cut &  Max-Cut  \\ 
\hline
 \textbf{Technology}  & CMOS  65nm GP & CMOS 65nm GP&Optical \& Digital & Optical  & CMOS 65nm GP & CMOS 65nm LP \\ 
 \hline
\textbf{Spins} & 2116 & 2000 & 47  & 30  & 2750 & 1968  \\ 
 \hline
 \textbf{Average power} & 283.4 mW & 1.548 W & DNR & DNR & 17.48 W &42 mW  \\ 
 \hline
\textbf{Time to solution} &  60 ns & 11 ns  & 500 \micro s & DNR  & 10 ns & 50 ns \\ 
 \hline
 \textbf{Accuracy} & 96\%-97\%* & 83\%-92\%* & 50\% success rate** & 50\%-100\%* &91\% - 94\%* & 89\%-100\%*   \\ 
 \hline
 \textbf{Baseline} & Exact solution & Exact Solution  & Exact solution & Exact solution &SA& Tabu \\ 
 \hline
\end{tabular}
 \begin{tablenotes}
            \small
             \item \raggedleft{* Accuracy is presented as the worst (if reported) and the best accuracy obtained w.r.t. the baseline over iterations}
            \item  \raggedleft{** Sucess rate refers to the number of times 100\% accuracy is achieved}
        \end{tablenotes}
 \end{threeparttable}
\end{table*}

\subsection{Performance Analysis on Different Problems }
\label{sec:analysis}

Four different custom-generated problems of different sizes are selected to study the variations in the performance of the MSROPM and the quality of solutions. King's graph topology graphs with 49, 400, 1024, and 2116 nodes with all edges active (8 edges per node) are mapped to the MSROPM. 40 iterations (i.e. repeated runs) are performed for each problem, allowing the MSROPM to explore the solution space. 

Figure~\ref{fig:plots}~(a) shows the 4-coloring accuracy obtained at each iteration with three of the four problems mapped to the MSROPM. For the smallest 4-coloring problem of 49 nodes, the MSROPM reaches 100\% accuracy (exact solution or global minimum) 6 times among 40 iterations. The average accuracy for the for the 49-node problem is 98\%, showing the consistency of the MSROPM at reaching high quality solutions even though worse solutions down to 92\% accuracy are sometimes obtained. For the 400-node problem, 97\% is the accuracy of the best solution. Due to the larger solution space, the range of accuracy that can be captured within 40 iterations is lower where the global minimum is never reached and the averaged accuracy of 97\% is slightly lower compared to the 49-node problem. 1024-node problem carries on this trend. Table~\ref{table:sol_size} shows the number of possible spin states (search space) for the 3 benchmark problems and a larger problem having 2116 nodes. The size of the solution space increases exponentially with the problem size, decreasing the likelihood of finding the global minimum. When the problem size is further increased to 2116 nodes, best solution accuracy stays at 97\%, with the continuation of the power scaling trend.

Since the MSROPM operates in 2 stages for the 4-coloring problem, it is important to study how the result of the $1^{st}$ stage (max-cut accuracy) affects the resulting 4-coloring accuracy in the $2^{nd}$ stage. Figure~\ref{fig:plots}~(b) demonstrates the max-cut accuracies obtained at each of the 40 iterations for the 3 problems. Looking at the accuracy of the two stages of the same iteration, results show that $1^{st}$ stage accuracy has, in general, positive correlation with the final 4-coloring accuracy, demonstrating the importance of the quality of results in both stages of solution.  

Hamming distances between the solutions obtained by the MSROPM are presented in the histograms of Figure~\ref{fig:plots}(c) as an indication of how different the solutions are from each other. 
The results show that solutions even with similar accuracies are significantly different (especially with increasing problem size), showing the randomness of the solutions and the probabilistic nature of the architecture. 

 Computation cycles of the MSROPM are allocated predetermined durations regardless of the problem size. OIM implementations reportedly have sub-linear or near-constant scaling for increasing problem-sizes through natural parallelization~\cite{wang_oim}. The random initialization of the ROSC phases at startup and between two stages is empirically set to last 5~ns until couplings are activated. The first (max-cut solving) and second (4-coloring solving) coupled annealing stage free of SHIL injection both last 20~ns which is empirically determined to be enough for the phases to reach nondiscretized, contended ground state where phases are not at fixed at targeted points. After the activation of SHIL\_1 in the first 2-phase discretization and both SHILs in the 4-phase discretization, 5~ns is allocated for stabilization and phase-readout. A complete run of the MSROPM lasts 60~ns in this scheme. The power consumption of the MSROPM implementation is below 283.4~mW, scaling linearly with increasing problem sizes.



\subsection{Comparison of Results with Prior Work}
Comparison with other Potts machines in the literature, implemented with similar and different technologies, and Ising machined implemented with similar (CMOS coupled ROSC and RTWO) technologies are presented in Table~\ref{table:comparison}. 

Reference \cite{potts_nature} solves a 47-node 4-coloring problem with a coherent Potts machine (CPM) with an additional digital component and multi-stage operation. Reference \cite{potts_nature} reports exact solutions with a 50\% success rate. The MSROPM converges to exact solutions with a similarly sized problem (49-node). In addition, MSROPM presents results with much larger problems up to 20 $\times$ the size of the 47-node problem. 
Reference~\cite{optical_potts} is another CPM implementation using optical parametric oscillators solving an 30-node 3-coloring problem. Reference~\cite{optical_potts} reports accuracy levels between 50\% and 100\% with an average of around 75\%, lower than the average 98\% accuracy achieved with the similarly sized problem (49-node) by MSROPM. Study~\cite{potts_nature} reports 500 \micro s run-time per stage, significantly higher than the 60~ns total run time of the MSROPM. Studies~\cite{optical_potts} do not report the time to solution and neither of these studies report power consumption. 

Reference \cite{ICCAD_potts} presents a similar ROSC-based Potts machine solving 3-coloring problems utilizing higher order SHIL. The Potts machine in \cite{ICCAD_potts} operates with a higher frequency of 7 GHz contributing to the higher power consumption and the faster solution times with comparable number of spins. Notwithstanding the difference in the solved COPs The accuracy of the Potts machine \cite{ICCAD_potts} is lower than the MSROPM showing the handicap of the N-SHIL method. The divide-and-conquer approach of the MSROPM allows reaching higher accuracies than the ROPM in \cite{ICCAD_potts} despite the increased difficulty of the 4-coloring problem over 3-coloring problem where the search spaces compare as $4^N$ to $3^N$ for similar problem sizes. 

The comparisons to the ROIM in~\cite{chris_kim_nat} and RTWOIM in~\cite{rtwo_ising} are performed cognizant of the differences and the similarities of the mapped COPs and architectures. The 4-coloring problem has a much larger number of possible energy states than max-cut i.e. $4^N$ to $2^N$, representative of the increased difficulty. In addition, the accuracies in~\cite{chris_kim_nat} and~\cite{rtwo_ising} are not with respect to the optimal solution. The power dissipation of the RTWOIM~[9] is orders of magnitude higher compared to the ROIM~[8] and the proposed MSROPM mainly due to the high frequency of operation of the RTWOIM. Consequently, the solution time of RTWOIM is 5$\times$ and 6$\times$ faster compared to ROIM and proposed MSROPM, respectively.  Comparing the similarly budgeted power profiles of the  ROIM~[8] and the proposed MSROPM, 
the relative increase in the power consumption is partially caused by the additional control circuitry, increased frequency, and the use of a different PDK. Lower operating frequency and multiple-stage operation of the MSROPM also lead to longer solution time compared to the RTWOIM and ROIM. 

\section {Conclusions}
\label{sec:conclusions}
This work presents a multi-stage ring-oscillator based Potts machine implementation obtaining multivalued spins from a single oscillator. Proposed MSROPM is able to solve 4-coloring problems with up to 100\% accuracy and down to only 97\% accuracy for the largest problem of 2116 nodes. The proposed MSROPM can be extended to solve COPs with more spin-values, with projected power, cost, and size advantages against its counterparts built with different technologies.



\bibliographystyle{IEEEtran}
\bibliography{ref}

\end{document}